\documentclass[reprint,prd,nofootinbib,aps]{revtex4-2}
\usepackage{amsmath,amssymb,amsfonts,graphicx,nicefrac,xcolor}
\usepackage{overpic}
\usepackage[T1]{fontenc}

\let\vrho\rho
\def\phi{\varphi}
\def\eps{\varepsilon}
\renewcommand{\rho}{\varrho}
\def\d{\mathrm{d}}
\def\p{\partial}

\def\arctanh{\mathop{\rm arctanh}}

\def\Re{\mathop{\rm Re}}
\def\Im{\mathop{\rm Im}}

\renewcommand{\vec}[1]{\boldsymbol{#1}}

\newcommand{\ka}{a}
\newcommand{\Rp}{r_{\!p}}
\newcommand{\Rm}{r_m}
\newcommand{\Krho}{\vrho}
\newcommand{\Krhocc}{\bar{\vrho}}

\newcommand{\lK}{\vec{\tilde{l}}}
\newcommand{\nK}{\vec{\tilde{n}}}
\newcommand{\mK}{\vec{\tilde{m}}}
\newcommand{\mKcc}{\vec{\bar{\tilde{m}}}}
\newcommand{\SigZ}{\Sigma_0}
\newcommand{\UpsPi}{\Upsilon_{\!\frac{\pi}{2}}}

\begin{document}
\title{Kerr black hole in the formalism of isolated horizons}
\author{David Kofro\v{n}}\email{d.kofron@gmail.com}
\affiliation{
Institute of Theoretical Physics, Faculty of Mathematics and Physics,\\
Charles University,\\
V Hole\v{s}ovi\v{c}k\'{a}ch 2, 180\,00 Prague 8, Czech Republic}


\begin{abstract}
We revise the work of Scholtz, Flandera and G\"urlebeck [Kerr-Newman black hole in the formalism of isolated horizons, Phys. Rev. D \textbf{96}, 064024 (2017)]. We cast the Kerr metric explicitly in the form suitable for the framework of isolated horizons. We proceed in a geometrical fashion and are capable to provide the results in a compact closed manner, without any unevaluated integrals. We also discuss the uniqueness and drawbacks of this construction. We suggest a new vector field to generate
the null geodesic foliation.
\end{abstract}
\maketitle

\section{Introduction}
The concept of weakly isolated horizons (WIH) has been developed at the beginning of this millennia by Ashtekar et al. \cite{ashtekar_isolated_1999,ashtekar_mechanics_2000,ashtekar_mechanics_2001,ashtekar_geometry_2002} in order to tackle the notion of a black hole in equilibrium state possibly surrounded by matter or radiation. Later on this concept was upgraded to notion of dynamical horizons \cite{ashtekar_isolated_2004} where even the general dynamical processes are taken into the account in full general relativity.

This formalism was used in \cite{krishnan_spacetime_2012} to evolve the solution from the initial data in a near-horizon expansion using the Newman\,--\,Penrose (NP) formalism \cite{Newman1962,penrose_spinors_nodate}.

The rotating Kerr black hole --- one of the most astrophysically relevant solution of the vacuum Einstein field equations --- was discovered in 1963 \cite{Kerr1963}. Recent historical reviews can be found in \cite{Wiltshire2009,Teukolsky2015}. There exists a plethora of interesting coordinate systems for the Kerr black hole and it is worth to investigate them on their own. Let us mention some of the most important 
\begin{itemize}
    \item Kerr\,--\,Schild form \cite{Kerr1963,chandrasekhar_mathematical_2009} in which it has been found,
    \item Boyer\,--\,Lindquist (BL) coordinates \cite{boyer_maximal_1967} which are one of the most suitable for interpretation and, simultaneously, the equations governing test fields are proven to be separable in these coordinates \cite{Teukolsky1973},
    \item Eddington\,--\,Finkelstein coordinates \cite{Kerr1963} which are regular on the horizon,
    \item Kruskal coordinates which reveals the full structure of the spacetime using the maximal analytical extension of the coordinates \cite{boyer_maximal_1967},
    \item Doran coordinates \cite{doran_new_2000} are adapted to family of time-like observers with zero angular momentum which are freely falling to the black hole,
    \item Bondi-like coordinates which are propagated from the null infinity \cite{fletcher_kerr_2003}.
\end{itemize}

At the same time the Kerr black hole itself is an excellent example of space-time containing a WIH. An attempt to cast it in this framework has been done in \cite{scholtz_kerr-newman_2017} in the recent years.

In this paper we will revise the work \cite{scholtz_kerr-newman_2017} which is difficult to follow and contains some implicit, unevaluated integrals; and, rather surprisingly, not a single picture of new coordinates is presented. Using a slightly different approach, we are able to evaluate all the integrals and derive simple results. In \cite{scholtz_kerr-newman_2017} the authors prescribe what the final vector field generating the null congruence should be and then they applied 4 successive transformations (boost, null rotation, spin, null rotation) on the Kinnersley tetrad to align the tetrad with this vector field. At each step they evaluated the NP spin coefficients and solve lot of equations in order to make some of he NP spin coefficients vanishing. Instead of this we simply construct a tetrad based directly on the final vector field, using the Killing\,--\,Yano tensor, and make just one null rotation so that the tetrad is parallely propagated. We also discuss the uniqueness of this construction and its drawbacks.

In the end, we reveal so many disadvantages of the null congruence chosen in \cite{fletcher_kerr_2003,scholtz_kerr-newman_2017} that we suggest a completely new congruence which is well behaved. Alas, this one is much more difficult to treat analytically.

The paper is organized as follows: we briefly introduce the concept of isolated horizons (IH) in Sec.~\ref{sec:iho} and recapitulate the the Kerr solution in Sec.~\ref{sec:Kerr}. In the Sec.~\ref{sec:pptetrad} we develop the parallely propagated null tetrad with the help of Killing\,--\,Yano tensors. The explicit formulae for the elliptic integral introduced in this section are postponed to Appendix~\ref{app:eli}. The null coordinates adapted to this tetrad are then introduced in Sec.~\ref{sec:null} and their affine parameterization in the following Sec.~\ref{sec:Aff}. The final form of the tetrad in these coordinates is contained in Sec.~\ref{sec:final}. In Sec.~\ref{sec:uni} we discuss the (non)uniqueness of this construction and propose an alternative solution. In Sec.~\ref{sec:bondi} we shortly compare these coordinates with the Bondi coordinates.

\section{Isolated horizons}
\label{sec:iho}
We follow the construction presented in \cite{krishnan_spacetime_2012,scholtz_kerr-newman_2017}\footnote{We had to adjust a sign on either $\vec{l}$ or $\vec{n}$ since we are using different signature of the metric.}, details can be found in \cite{ashtekar_geometry_2002}. 

On the manifold $M$ the \emph{non-expanding horizon} $\mathcal{H}\subset M$ is defined as a null hypersurface with following properties
\begin{itemize}
    \item the topology of $\mathcal{H}$ is $R\times S^2$
    \item the expansion of any null normal $\vec{l}^a$ is vanishing
    \item equations of motion are satisfied on $\mathcal{H}$ and the projection of stress energy tensor $\vec{T}_{ab}$ on any future pointing null normal $\vec{l}$ is causal (a consequence of dominant energy condition) 
\end{itemize}

It is not possible to define a unique induced covariant derivative on a general null hypersurface, since the induced metric is degenerate. However, the existence of additional conditions (namely the vanishing expansion of null normals) on the non-expanding horizon allows us to define a preferred connection $\mathcal{D}_a$ on the tangent bundle $T\mathcal{H}$ as
\begin{align}
    \vec{X}^a \mathcal{D}_a \vec{Y}^b &\doteq \vec{X}^a \nabla_a \vec{Y}^b\,, \quad \text{for any }\vec{X}^a,\,\vec{Y}^a\,\in T\mathcal{H}\,,
\end{align}
where $\nabla_a$ is connection on $TM$ compatible with the metric $\vec{g}_{ab}$ and the relation operator ``$\doteq$'' means that the equality holds on $\mathcal{H}$. 

This covariant derivative implies the existence of a \emph{rotational 1-form} $\vec{\omega}$ which is defined by
\begin{align}
    \mathcal{D}_a \vec{l}^b &\doteq \vec{\omega}_a\vec{l}^b\,.
\end{align}

The null normal $\vec{l}$ can be rescaled by an arbitrary non-vanishing function. In order to get rid of this ambiguity we define the \emph{weakly isolated horizon} as a non-expanding horizon $\mathcal{H}$ together with the equivalence class of null normals $[\vec{l}]=\{c \vec{l},\,c\in \mathbb{R},\,c\neq 0\}$ (the remaining freedom is scaling by constant) and restrict to the class for which a representative element $\vec{l}\in[\vec{l}]$ satisfies
\begin{align}
    \left[\mathcal{L}_{\vec{l}},\mathcal{D}_a\right]\vec{l}^b &\doteq 0\,, \label{eq:nullL}
\end{align}
where $\mathcal{L}$ is the Lie derivative. This condition is equivalent to $\mathcal{L}_{\vec{l}} \vec{\omega}_a\doteq 0$.

The weakly isolated horizons are physically identical to non-expanding horizons. The selection of the proper equivalence class of null normals $[\vec{l}^a]$, given by Eq.~(\ref{eq:nullL}) allowed to formulate zeroth law of thermodynamics (which is equivalent to Eq.~(\ref{eq:nullL})) for isolated horizons and give a physical meaning for the rotational 1-form $\vec{\omega}$, see \cite{ashtekar_mechanics_2001,krishnan_spacetime_2012}.

Choosing a particular null generator $\vec{l}$ of a WIH $(\mathcal{H},[\vec{l}])$ we can (non-uniquely) complete the full NP tetrad $(\vec{l},\,\vec{n},\,\vec{m},\,\vec{\bar{m}})$ so that $\vec{m},\vec{\bar{m}} \in T\mathcal{H}$ and $\vec{n}$ is transversal to $\mathcal{H}$. 

Having the NP tetrad we can demonstrate the significance of rotational 1-form. It has been proven in \cite{ashtekar_geometry_2002} that
\begin{align}
    \d \vec{\omega} \doteq \Im \left(\Psi_2\right) \vec{\varepsilon}\,,
\end{align}
where $\vec{\varepsilon}$ is a volume element of cross sections of $\mathcal{H}$ and $\Psi_2$ is the standard projection of Weyl tensor.

Our primary goal is to set up a coordinate system adapted to the isolated horizon and NP formalism (an analog of Bondi system).

The null generator $\vec{l}$ gives rise to a preferred foliation $\mathcal{S}_v$ of $\mathcal{H}$ by topological spheres by imposing $D v\equiv \vec{l}^a\nabla_a v \doteq 1$. On the spherical cut $\mathcal{S}_0$ of $\mathcal{H}$ angular coordinates $\zeta^2,\zeta^3$ are introduced and they are propagated on $\mathcal{H}$ along the null generators, i.e. we have $D \zeta^i\doteq 0$ for $i=(2,3)$. The vectors $\vec{m},\,\vec{\bar{m}}$ are Lie dragged along the horizon as $\mathcal{L}_{\vec{l}} \vec{m}^a \doteq \mathcal{L}_{\vec{l}} \vec{\bar{m}}^a \doteq 0$.

Finally, the null tetrad defined on the horizon is parallely propagated outward in the transversal direction defined by $\vec{n}$. This null geodetic congruence is affinely parameterized by $s$. Starting from a point on the horizon $(v,\zeta^2,\zeta^3)$ every point at least in the vicinity of $\mathcal{H}$ should be assigned  coordinates $(v,s,\zeta^2,\zeta^3)$ by this procedure (for the eternal Kerr black hole we want to construct a global coordinate system; for more complicated solutions the construction of the coordinates is not guaranteed globally).

Thus we look for a coordinate system $(v,s,\Theta,\tilde{\Phi})$\footnote{We coined a particular names to $\zeta^i$, namely $\zeta^2=\Theta$ and $\zeta^3=\tilde{\Phi}$.} and a null tetrad such that the metric
\begin{align}
    \vec{g}_{ab}=2\vec{l}_{(a}\vec{n}_{b)}-2\vec{m}_{(a}\vec{\bar{m}}_{b)} \,,
\end{align}
is reconstructed from a NP tetrad in the form
\begin{align}
    \vec{l} &= \vec{\p_v}+U\vec{\p_s}+X^\Theta\vec{\p_\Theta}+X^{\tilde{\Phi}}\vec{\p_{\tilde{\Phi}}}\,, \\
    \vec{n} &= \vec{\p_s}\,,\\
    \vec{m} &= \Omega\vec{\p_s}+\xi^\Theta\vec{\p_\Theta}+\xi^{\tilde{\Phi}}\vec{\p_{\tilde{\Phi}}}\,,
\end{align}
satisfying following conditions on the horizon: (a) vector $\vec{l}$ is generator of the horizon, (b) vectors $\vec{m},\vec{\bar{m}}$ are tangent to the horizon, (c) the vector field $\vec{n}$ is transversal to the horizon, geodetic, affinely parameterized and twist-free. The whole tetrad is parallely propagated along $\vec{n}$. 

In NP formalism these conditions translates to the following conditions on the NP scalars
\begin{equation}
    \gamma = \tau = \nu = \mu-\bar{\mu} = \pi-\alpha-\bar{\beta} =0\,,
\end{equation}
which hold everywhere and 
\begin{equation}
    \rho \doteq \kappa \doteq \sigma \doteq \epsilon-\bar{\epsilon} \doteq 0\,,
\end{equation}
and
\begin{equation}
    U\doteq X^\Theta \doteq X^{\tilde{\Phi}} \doteq \Omega \doteq 0\,,
\end{equation}
which hold on the horizon $\mathcal{H}$ only.

\section{Kerr black hole}
\label{sec:Kerr}

The standard form of the Kerr metric in BL coordinates \cite{boyer_maximal_1967} reads 
\begin{multline}
\vec{\d} s^2 = -\frac{\Delta}{\Sigma} \left( \vec{\d} t -\ka\sin^2\theta\,\vec{\d}\phi \right)^2 +\frac{\Sigma}{\Delta}\,\vec{\d} r^2 \\
+ \Sigma \, \vec{\d}\theta^2  
+\frac{\sin^2\theta}{\Sigma}\Bigl( \left( \ka^2+r^2 \right)\vec{\d}\phi - \ka\,\vec{\d} t \Bigr)^2\,,
\label{eq:KerrMetric}
\end{multline}
with the definitions 
\begin{align}
\Delta(r)&=r^2-2Mr+\ka^2=(r-r_p)(r-r_m)\,,\\
\vrho(r,\theta)&=r-i\ka\cos\theta\,,\\
\Sigma(r,\theta)&=\vrho\bar{\vrho}=r^2+\ka^2\cos^2\theta\,.
\end{align}
The parameters have the following meaning: $M$ is the mass of the black hole, $Ma$ is its angular momentum, $r_p$ is the position of outer black hole horizon whereas $r_m$ is the position of inner black hole horizon. We will use all of these parameters ($r_p,r_m,M,a$) although only two of them are independent.

The Kerr metric is endowed by two Killing vector fields, namely, the one associated with stationarity $\vec{\xi}=\vec{\p_t}$ and the one associated with axial symmetry $\vec{\eta}=\vec{\p_\phi}$. It is advantageous to introduce also a quantity
\begin{equation}
\Upsilon(r,\theta) = -\frac{\Sigma}{\sin^2\theta}\,\vec{\eta}\cdot\vec{\eta}=\Delta\Sigma+2Mr\left(r^2+a^2\right).
\end{equation}

The Kinnersley NP\footnote{Notice the boost given by $\sqrt{2}$ in contrast to the standard textbook form. The Kinnersley tetrad is denoted by tilde in this work.} tetrad  $(\vec{\tilde{l}},\,\vec{\tilde{m}},\,\vec{\tilde{\bar{m}}},\,\vec{\tilde{n}})$, introduced in \cite{kinnersley_type_1969}, which is adapted to the principal null directions of the Weyl tensor reads as follows
\begin{equation}
\begin{alignedat}{1}
\vec{\tilde{l}} &= \frac{1}{\sqrt{2}\,\Delta}\left[ \left( r^2+\ka^2 \right)\vec{\p_t}+\Delta\,\vec{\p_r} + \ka\,\vec{\p_\phi} \right]\,, \\
\vec{\tilde{n}} &= \frac{1}{\sqrt{2}\,\Sigma}\left[ \left( r^2+\ka^2 \right)\vec{\p_t}-\Delta\,\vec{\p_r} + \ka\,\vec{\p_\phi} \right]\,, \\
\vec{\tilde{m}} &= \frac{1}{\sqrt{2}\,\bar{\vrho}}\bigl( i\ka\sin\theta\,\vec{\p_t}+\vec{\p_\theta} +i\csc\theta\,\vec{\p_\phi} \bigr)\,.
\end{alignedat}
\label{eq:NPtetrad}
\end{equation}
The appropriate spin coefficients are enlisted in Appendix~\ref{app:Kerr}.

The Kerr metric posses not only two Killing vector fields but also a principal Killing\,--\,Yano  tensor \cite{kalnins_killingyano_1989,Kubiznak2007} which provides additional hidden symmetry. The principal Killing\,--\,Yano tensor expanded in the Kinnersley tetrad is given by
\begin{align}
    \vec{f}_{ab} &=-2ir\,\mK_{[a}\mKcc_{b]}+2\ka\cos\theta\,\lK_{[a}\nK_{b]}\,,
\end{align}
and its dual is also a Killing\,--\,Yano tensor, namely
\begin{align}
    \vec{h}_{ab} &=2r\,\lK_{[a}\nK_{b]}+2i\ka\cos\theta\, \mK_{[a}\mKcc_{b]}\,.
\end{align}

From now on, we use the following shortcuts, to make the formulae visually more compact
\begin{align}
    \SigZ &= \Sigma(r,0)=r^2+\ka^2\,, &
    \UpsPi &= \Upsilon(r,\pi/2) \,.
\end{align}

Using the four independent constants of motion we can write a general geodesic congruence as follows
\begin{align}
\vec{u} &= \frac{1}{\Sigma}\Biggl[
\left( \frac{\SigZ\left(\SigZ E-\ka L\right)}{\Delta}+\ka\left(L-a\sin^2\theta\,E\right)\right)\vec{\p_t} \nonumber\\
& \phantom{=} + \epsilon_r \sqrt{\left(\SigZ E-\ka L\right)^2-\left(K-\kappa r^2\right)\Delta}\,\vec{\p_r} \nonumber\\
& \phantom{=} + \epsilon_\theta\sqrt{K+\kappa \ka^2\cos^2\theta-\csc^2\theta\left(L-a\sin^2\theta E\right)} \,\vec{\p_\theta} \nonumber\\
& \phantom{=} + \left(\csc^2\theta\left(L-\ka\sin^2\theta E\right)^2+\frac{\ka\left(\SigZ E-aL\right)}{\Delta}\right) \vec{\p_\phi} 
\Biggr].
\label{eq:genu}
\end{align}
The parameters $E$ (energy), $L$ (angular momentum) and $K$ (Carter's constant) are constants along a single geodesic. They can be functions of $(r,\theta)$, however, they have to obey 
\begin{align}
    \vec{u}^a\nabla_a E &= 0\,, &
    \vec{u}^a\nabla_a L &= 0\,, &
    \vec{u}^a\nabla_a K &= 0\,,
\end{align}
for $\vec{u}$ being the geodesic congruence.

Let us note that Doran coordinates are adapted to the geodetic congruence $\vec{u}$ with the constants chosen as $\kappa=-1,\,L=0,\,E=-1,\epsilon_r=1$ and are independent on the value of $\epsilon_\theta$ (since the $\theta$ component vanishes).

\section{The parallely propagated NP tetrad}
\label{sec:pptetrad}

In order to obtain a suitable null foliation of the Kerr spacetime we look for a twist-free null geodesic congruence, which penetrates horizon and can reach the null infinity. To cover the regions close to the axis, the angular momentum has to be zero (as can be seen from the $\theta$ component of $\vec{u}$). One of the possibilities\footnote{The choice $\epsilon_\theta=1$ was discussed in \cite{scholtz_kerr-newman_2017} and we will cover it in Appendix~\ref{sec:differentEps}.} is to demand 
\begin{align}
    E &= -1\,, &
    L &= 0\,, &
    \kappa &= 0 \,, &
    K &= \ka^2 \,, \label{eq:const1}\\
     &&
    \epsilon_r &= 1\,,&
    \epsilon_\theta &= -1\,, && \label{eq:const2}
\end{align}
which also ensures the mirror symmetry with respect to the equator.

The vector $\vec{u}$, given in Eq.~(\ref{eq:genu}), with constants of motion given by Eqs.~(\ref{eq:const1}\,--\,\ref{eq:const2}) then becomes the first vector of our NP tetrad
\begin{align}
\vec{n} &= \frac{-1}{\Sigma}\Bigg(
\frac{\Upsilon}{\Delta}\,\vec{\p_t}-\sqrt{\UpsPi}\,\vec{\p_r}+\ka\cos\theta\,\vec{\p_\theta}+\frac{2\ka M r}{\Delta}\,\vec{\p_\phi}
\Biggr),
\label{eq:n}
\end{align}
since
\begin{align}
\vec{n}^a\nabla_a\vec{n}^b &=0\,.
\end{align}

Such a geodetic congruence is generated by null rays passing through the horizon and reaching the null infinity. We may introduce angular coordinates $\Theta$ and $\Phi$ which will be constant along these null rays, namely
\begin{align}
    \Theta &= 2\arctan\left[\tanh \Gamma^+(r,\theta)\right]\,, \label{eq:Theta}\\
    \Phi &= \phi+2aM\int_{r_p}^r\frac{u}{\Delta(u)\sqrt{\UpsPi(u)}}\,\d u\,,\label{eq:Phi}
\end{align}
where
\begin{align}
    \Gamma^+(r,\theta) & = \arctanh\left[\tan\left(\frac{\theta}{2}\right)+\frac{\ka}{2}\mathcal{I}_0(r) \right]\,.\label{eq:gammaP} 
\end{align}
Here we introduced the integral $\mathcal{I}_1$ (and also $\mathcal{I}_2$ for further reference) as
\begin{align}
    \mathcal{I}_0(r) &= \int_{r_p}^r\frac{1}{\sqrt{\UpsPi(u)}}\,\d u \,, \label{eq:I0}\\
    \mathcal{I}_2(r) &= \int_{r_p}^r\frac{u^2}{\sqrt{\UpsPi(u)}}\,\d u \,. \label{eq:I2}
\end{align}
These integrals are explicitly given in terms of elliptic integrals -- see Appendix~\ref{app:eli} for particular formulae.

Thus, we have
\begin{align}
    \vec{n}^a\nabla_a \Theta(r,\theta) &= 0\,, &
    \vec{n}^a\nabla_a \Phi(r,\theta) &= 0\,,
\end{align}

The inverse of the relations (\ref{eq:Theta})\,--\,(\ref{eq:Phi}), i.e. how $\theta$ and $\phi$ change along the null ray defined by $\Theta=\mathrm{const}$ and $\Phi=\mathrm{const}$, read
\begin{align}
    \theta &= 2\arctan\left[
    \tanh \Gamma^-(r,\Theta) \right]\,, \label{eq:theta}\\
    \phi &= \Phi-2aM\int_{r_p}^r\frac{u}{\Delta(u)\sqrt{\UpsPi(u)}}\,\d u\,,\label{eq:phi}
\end{align}
where
\begin{align}
    \Gamma^-(r,\Theta) &=  \arctanh\left(\tan\frac{\Theta}{2}\right)-\frac{a}{2}\,\mathcal{I}_0(r) \,.
\end{align}

The program of the construction of the tetrad and coordinate system is now straightforward
\begin{enumerate}
    \item Employing the Killing\,--\,Yano tensors we complete the NP tetrad\footnote{The dots over the vectors does not represent derivatives. We choose them instead of indices $(1)$ and $(2)$ which would spoil the equations.} to get  $\left(\vec{\ddot{n}}=\vec{n},\,\vec{\ddot{l}},\,\vec{\ddot{m}},\,\vec{\ddot{\bar{m}}}\right)$. See the explicit construction below in Eqs.~(\ref{eq:ptky}\,--\,\ref{eq:tetrad1}).
    \item We perform null rotation around $\vec{n}$, so that the all the vectors are parallely propagated along $\vec{n}$ and, moreover, on the horizon $\vec{l}$ coincides with its generator, to obtain $\left(\vec{n},\,\vec{l},\,\vec{\dot{m}},\,\vec{\dot{\bar{m}}}\right)$. This is done in Eqs.~(\ref{eq:nullrot})\,--\,(\ref{eq:rotE}).
    \item If desired, we may perform final rotation in $\vec{\dot{m}}-\vec{\dot{\bar{m}}}$ plane (using functions of $\Theta$ coordinate, such that the parallel transport of the new vector is not violated) and thus obtain $\left(\vec{n},\,\vec{l},\,\vec{m},\,\vec{\bar{m}}\right)$.
    \item Finally, we perform the coordinate transformation to the coordinates of the isolated horizon formalism, see Sections~\ref{sec:null},\ref{sec:Aff}.
\end{enumerate}

A well known fact is that the principal Killing\,--\,Yano tensor can be used to produce a parallely transported vector along the geodesic congruence given by $\vec{n}$
\begin{align}
    \vec{e_{(f)}}^b &= \frac{1}{\ka}\,\vec{f}^{b}_{\phantom{b}c}\vec{n}^{c}\,, \label{eq:ptky}
\end{align}
which is space-like in our case. We can complete the tetrad by the following two vectors:
\begin{align}
    \vec{e_{(h)}}^b &= \frac{1}{\ka}\,\vec{h}^{b}_{\phantom{b}c}\vec{n}^c\,,\\
    \vec{{e_{(3)}}}^b &= \frac{1}{\ka^2}\left(\vec{f}^b_{\phantom{b}c} 
    \vec{f}^c_{\phantom{c}d}\vec{n}^d+\vec{f}_{jk}\vec{f}^{jk} \vec{n}^b\right).
\end{align}
From these we can construct a null tetrad as
\begin{multline}
    \left(\vec{n},\,\vec{\ddot{l}},\,\vec{\ddot{m}},\,\vec{\ddot{\bar{m}}}\right) \\
    =\left(
    \vec{n},\, 
    \vec{e_{(3)}},\,
    \frac{1}{\sqrt{2}}\left(\vec{e_{(f)}}+i\vec{e_{(h)}}\right),
    \frac{1}{\sqrt{2}}\left(\vec{e_{(f)}}-i\vec{e_{(h)}}\right)
    \right),\label{eq:tetrad1}
\end{multline}
which explicitly reads as follows
\begin{align}
\vec{n} &=\mathrm{Eq.\ (\ref{eq:n})}\,, \\
\vec{\ddot{l}} &= -\frac{1}{2}\Biggl(
    \frac{\Sigma_0^2+\ka^2\sin^2\theta\,\Delta}{\ka^2\Delta}\,\vec{\p_t}
    -\frac{\sqrt{\UpsPi}}{\ka^2}\,\vec{\p_r} \nonumber\\
    &\qquad\qquad
    -\frac{\cos\theta}{\ka}\,\vec{\p_\theta}
    +\frac{\Sigma_0+\Delta}{\ka\Delta}\,\vec{\p_\phi}\Biggr), \\
\vec{\ddot{m}} &= -\frac{1}{\sqrt{2}\bar{\vrho}}\Biggl(
    \frac{\ka^2\sin 2\theta\, \Delta+i\Sigma_0\sqrt{\UpsPi}}{2\ka\Delta}\,\vec{\p_t}
    -\frac{i\Sigma_0}{\ka}\,\vec{\p_r} \nonumber\\
    &\qquad\qquad +\sin\theta\,\vec{\p_\theta}
    +\frac{\cot\theta\,\Delta+i\sqrt{\UpsPi}}{\Delta}\,\vec{\p_\phi}\Biggr).
\end{align}
The Weyl scalars in this tetrad have the following form 
\begin{align}
    \left(\ddot{\psi}_0,\ddot{\psi}_1,\ddot{\psi}_2,\ddot{\psi}_3,\ddot{\psi}_4\right) =-\left(
    \frac{3}{4}\frac{\vrho^2}{\ka^2}\,,
    0\,,
    \frac{1}{2}\,,
    0\,,\frac{3}{4}\,
    \frac{\ka^2}{\vrho^2}\right)\tilde{\psi}_2\,,
\end{align}
where $\tilde{\psi}_2$ is the only non-vanishing Weyl projection in the Kinnersley tetrad (\ref{eq:NPtetrad}), which is adapted to principal null directions of the Weyl tensor.

The appropriate spin coefficients are as follows
\begin{align}
    \ddot{\kappa} &= \frac{-\Sigma_0 \cos\theta +i\ka r\sin^2\theta}{\sqrt{2}\ka^2\bar{\vrho}} \,, \\
    \ddot{\nu} &= 0 \,, \\
    \ddot{\sigma} &=\frac{1}{16\ka^2\bar{\vrho}}\left(4\ka\frac{i\ka\cos\theta-r\cos 2\theta}{\sin\theta}+\frac{\ddot{\upsilon}}{\bar{\vrho}\sqrt{\UpsPi}}\right) \,,\\
    \ddot{\lambda} &= \frac{1}{8\vrho^3}\left(4\ka\frac{i\ka\cos\theta-r\cos 2\theta}{\sin\theta}-\frac{\ddot{\upsilon}}{\bar{\vrho}\sqrt{\UpsPi}}\right) \,,\\
    \ddot{\rho} &= \frac{1}{8\ka^2}\left(-2\ka\frac{\cos2\theta}{\sin\theta}-\frac{\p_r \UpsPi}{\sqrt{\UpsPi}}\right), \\
    \ddot{\mu} &= \frac{1}{4\Sigma}\left(-2\ka\frac{\cos2\theta}{\sin\theta}+\frac{\p_r \UpsPi}{\sqrt{\UpsPi}}\right) \,, \\
    \ddot{\tau} &= -\frac{i}{\sqrt{2}\ka} \,,\\
    \ddot{\pi} &= -\frac{i}{\sqrt{2}\ka} \,,\\
    \ddot{\beta} &= -\frac{i}{\sqrt{2}\ka} \,,\\
    \ddot{\alpha} &= 0\,, \\
    \ddot{\epsilon} &= \frac{i\ka^2\cos\theta\,\sin\theta+\sqrt{\UpsPi}}{2\ka^2\vrho}\,,\\
    \ddot{\gamma} &= 0\,,
\end{align}
where we defined a shortcut
\begin{align}
    \ddot{\upsilon} &= -8i\ka\UpsPi\,\cos\theta \\
    &\qquad +\ka^2\left(2M\ka^2+2\ka^2r-12Mr^2+\cos 2\theta\,\p_r\UpsPi\right).\nonumber
\end{align}

Notice that, at this stage, the limit $\ka\rightarrow 0$ does not exist for the tetrad $\left(\vec{n},\,\vec{\ddot{l}},\,\vec{\ddot{m}},\,\vec{\ddot{\bar{m}}}\right)$. However, the limit will be recovered once the final tetrad is constructed.

The vector $\vec{n}$ is already fixed, we thus perform a null rotation about it to get a new tetrad, namely
\begin{multline}
    \left(\vec{n},\,\vec{l},\,\vec{\dot{m}},\,\vec{\dot{\bar{m}}}\right)\\
    =
    \left(\vec{n},\,
    \vec{\ddot{l}}+E\vec{{\ddot{\bar{m}}}}+\bar{E}\vec{\ddot{m}}+E\bar{E}\vec{n},\,
    \vec{\ddot{m}}+E\vec{n},\,
    \vec{\ddot{\bar{m}}}+\bar{E}\vec{n}\right).\label{eq:nullrot}
\end{multline}
Imposing the condition of parallel transport of vector $\vec{\dot{m}}$ along $\vec{n}$, i.e. demanding
\begin{align}
    \vec{n}^a\nabla_a\vec{\dot{m}} &= 0\,,
\end{align}
leads to a first order partial differential equation for $E$. Its real and imaginary parts are as follows
\begin{align}
    \left(-\sqrt{\UpsPi}\,\p_{r}+a\cos\theta\,\p_\theta\right) \Re(E) &=0\,, \\
    \left(\Sigma-\sqrt{2}\ka\sqrt{\UpsPi}\,\p_r+\sqrt{2}\ka^2\cos\theta\,\p_\theta \right) \Im(E) &= 0\,.
\end{align}
We may try to solve for $E$ in a separated form $E(r,\theta)=E_r(r)+E_\theta(\theta)$. The general separable solution reads 
\begin{align}    
    E &=
    i\frac{r_p-\ka\sin\theta+\mathcal{I}_2}{\sqrt{2}\ka} +F(\Gamma^+)\,,
\end{align}
where $F$ is an arbitrary complex function of $\Gamma^+(r,\theta)$ which is defined in Eq.~(\ref{eq:gammaP}).

If $\vec{n}$ and $\vec{\dot{m}}$ (and thus also $\vec{\dot{\bar{m}}}$) are parallely transported along $\vec{n}$ then, inevitably, also the last vector $\vec{\dot{l}}$ is parallely transported.

This freedom in the choice of $F$ allows us to fix $\vec{l}$ in such a way that on the horizon it coincides with the generator of the horizon. The parallely transported tetrad is given by the function
\begin{align}    
    E &=\frac{1}{\sqrt{2}} \left(
    i\frac{r_p-\ka\sin\theta+\mathcal{I}_2}{\ka}  -\sec 2\Gamma^++i\tanh 2\Gamma^+
    \right).
\label{eq:rotE}
\end{align}

The Weyl scalars in this tetrad are related to the only non-vanishing projection $\tilde{\psi}_2$ of the Weyl tensor onto the Kinnersley tetrad  as follows
\begin{align}
    \dot{\psi}_0 &= -\frac{3}{4}\left(\frac{2\ka^2 \bar{E}^2+\vrho^2}{\ka\vrho}\right)^2\tilde{\psi}_2 \,, \\
    \dot{\psi}_1 &= -\frac{3}{2}\,\frac{2\ka^2\bar{E}^2+\vrho^2\bar{E}}{\vrho^2}\,\tilde{\psi}_2 \,, \\
    \dot{\psi}_2 &= -\frac{1}{2}\frac{6\ka^2\bar{E}^2+\vrho^2}{\vrho^2}\,\tilde{\psi}_2 \,, \\
    \dot{\psi}_3 &= -\frac{3\ka^2\bar{E}}{\vrho^2}\,\tilde{\psi}_2 \,, \\
    \dot{\psi}_4 &= -\frac{3\ka^2}{\vrho^2}\,\tilde{\psi}_2\,.
\end{align}

At this point we need to investigate the behaviour of the Weyl scalars (and of the null tetrad itself) in the limit $\ka\rightarrow 0$. Careful calculations show that 
\begin{equation}
    \bar{E}\approx -i\frac{r}{\sqrt{2}}\frac{1}{\ka}-\frac{\cos\theta}{\sqrt{2}}+\mathcal{O}(\ka)\,,
\end{equation}
and thus
\begin{align}
    \dot{\psi}_0 &= \mathcal{O}(\ka^2)\,,\\
    \dot{\psi}_1 &= \mathcal{O}(\ka)\,,\\
    \dot{\psi}_2 &= \bigl.\tilde{\psi}_2\bigr|_{\ka=0}+\mathcal{O}(\ka)\,,\\
    \dot{\psi}_3 &= \mathcal{O}(\ka)\,,\\
    \dot{\psi}_4 &= \mathcal{O}(\ka^2)\,.
\end{align}

The NP spin coefficients of the tetrad $(\vec{n},\vec{l},\vec{\dot{m}},\vec{\dot{\bar{m}}})$ are given by standard transformation rules (see \cite{Newman1962,penrose_spinors_nodate} for details).

The last, optional, step is to perform a rotation in $\vec{\dot{m}}-\vec{\dot{\bar{m}}}$ plane
\begin{equation}
    \left(\vec{n},\,\vec{l},\,\vec{m},\,\vec{\bar{m}}\right)
    =
    \left(\vec{n},\,
    \vec{l},\,
    e^{i\alpha}\vec{\dot{m}},\,
    e^{-i\alpha}\vec{\dot{\bar{m}}}\right),
\end{equation}
so that $\vec{e}_\phi=-i(\vec{m}-\vec{\bar{m}})/\sqrt{2}$ has only $t$ and $\phi$ components on the horizon. This can be done with
\begin{align}
    \alpha &= \arctan\left(\frac{\ka \cos\Theta}{r_p}\right)-\Theta-\frac{\pi}{2}\,.
\end{align}
This way we also reconstruct the non-rotating limit
\begin{align}
    \lim_{\ka\rightarrow 0} \vec{m} = \lim_{\ka\rightarrow 0} \vec{\tilde{m}} \,,
\end{align}
and the Weyl scalars are affected in a trivial way.

\section{Null coordinates}
\label{sec:null}
Following \cite{fletcher_kerr_2003,scholtz_kerr-newman_2017}, we define ingoing null coordinates $(v,r,\Theta,\Phi)$ by the relations
\begin{align}
    t &= -v -\int_{r_p}^r\frac{\sqrt{\UpsPi(u)}}{\Delta(u)}\,\d u +\ka \sin \theta(r,\Theta) \,, \label{eq:BLtoINt}\\
    r &= r \,, \\
    \theta &= 2\arctan\left[
    \tanh\Gamma^-(r,\Theta)\right]\,, \label{eq:BLtoINtheta} \\
    \phi &= \Phi-2aM\int_{r_p}^r\frac{u}{\Delta(u)\sqrt{\UpsPi(u)}}\,\d u\,,
\end{align}
where $\theta(r,\Theta)$ entering (\ref{eq:BLtoINt}) is defined in Eq.~(\ref{eq:BLtoINtheta}). In this coordinates the vector $\vec{n}$
takes the form\footnote{We did not want to rename all the coordinates and we thus write the vanishing components explicitly in gray color. Compare the form of $\vec{n}$ given in Eq.~(\ref{eq:n}) and Eq.~(\ref{eq:nn}).}
\begin{align}
\vec{n} &= {\color{lightgray}0\, \vec{\p_v}+} 
            \frac{\sqrt{\UpsPi(r)}}{\Sigma(r,\theta(r,\Theta))}\,\vec{\p_r} 
           {\color{lightgray}+0\, \vec{\p_\Theta}+0\,\vec{\p_\Phi}}\,. \label{eq:nn}
\end{align}

\section{Affinely parameterized null geodesics}
\label{sec:Aff}
The geodesic on Kerr background has been shown to be separable in Mino time \cite{Mino2003}. Yet, we need an affine parameter, not a Mino time.
It may seem hopeless to evaluate the integral which enters the definition of an affine parameter for the null ray given by $\vec{n}$
\begin{equation}
    s(r,\Theta)=\int_{r_p}^r \frac{\Sigma(u,\theta(u,\Theta))}{\sqrt{\UpsPi(u)}}\,\d u \,,
\end{equation}
but we found that the integral can be analytically calculated and the result is
\begin{align}
    s(r,\Theta)
    &=\ka\left[\sin\Theta- \tanh\Gamma^-(r,\Theta)\right]+\mathcal{I}_2(r) \nonumber\\
    &=\ka\left[\sin\Theta-\sin\theta(r,\Theta)\right]+\mathcal{I}_2(r) \,,
    \label{eq:sint}
\end{align}
where we use the function $\mathcal{I}_2$ which has been defined in Eq.~(\ref{eq:I2}).

The very last step would have been to invert this relation to get $r=r(s,\Theta)$ in order to be able to explicitly express the final expressions in terms of $v,s,\Theta,\tilde{\Phi}$  (let us remind that along a particular given ray $\Theta$ is constant). Alas, we were unable to proceed in this direction.

In order to eliminate the axial component of the vector $\vec{\dot{l}}$ on the horizon, we also perform a final coordinate transformation to coordinates $(v,s,\Theta,\tilde{\Phi})$ defined by
\begin{align}
v &= v\,, \\
s &= s(r,\Theta) \,,\\
\Theta &= \Theta \,, \\
\tilde{\Phi} &= \Phi + \frac{\ka}{\ka^2+\Rp^2}\,v \,,
\end{align}
where $\ka/(\ka^2+\Rp^2)$ is the angular velocity of the horizon. The coordinate $s$ is now the affine parameter along the null geodesics given by $\vec{n}$. 

\begin{figure}
    \centering
    \begin{overpic}[keepaspectratio,width=.2\textwidth,percent]{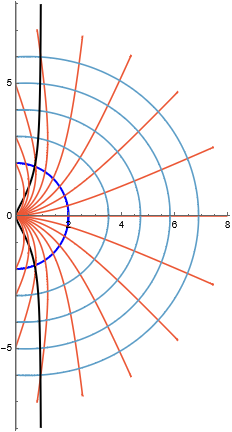}
    \put(55,42){$R$}
    \put(-2,95){$z$}
    \end{overpic}
    \caption{Red lines represent projection of integral lines of $\vec{n}$ (i.e. lines of constant $\Theta$) in BL coordinates meanwhile the light blue lines are equipotentials of constant affine parameter $s(r,\theta)$. The thick blue half-circle is black hole horizon $r=\Rp$, axis of rotation is vertical. The black line denotes vanishing expansion of the congruence $\vec{n}$. 
    \newline
    Calculations are done in BL coordinates, which are later identified with polar coordinates $(R=r\sin\theta,\,z=r\cos\theta$).}
    \label{fig:coords}
\end{figure}


The newly constructed coordinates $(s,\Theta)$ are visualised in Fig.~\ref{fig:coords}. The coordinates $\Theta$ and $\theta$ coincide on the outer horizon.

The null geodesic vector field $\vec{n}$ is singular near the axis (has non-vanishing $\theta$ component). This has some unpleasant consequences, as the presence of caustics.

From the definition of the transformation $\theta=\theta(r,\Theta)$ we can see that the range of variable $\Theta$ is $r$-dependent
\begin{align}
    \Theta &\in \langle \Theta_0(r),\,\pi-\Theta_0(r) \rangle\,, \\
    \Theta_0(r) &= 2\arctan\tanh\left(\frac{1}{2}\,\ka\,\mathcal{I}_0(r)\right)\,,
\end{align}
and some regions below the horizon are not covered at all.

It is peculiar that the expansion, i.e., an invariant quantity, of null geodetic congruence $\vec{n}$,
\begin{equation}
    \Theta_{\vec{n}} = \nabla^a\vec{n}_a = \frac{1}{\Sigma}\left(\frac{\d \UpsPi(r)}{\d r}-\ka\,\frac{\cos 2\theta}{\cos\theta} \right) ,
\end{equation}
is divergent close to the axis (as visualised in Fig.~\ref{fig:coords}).

\section{Final tetrad and coordinate system}
\label{sec:final}
In the coordinates constructed in the last section we can write down the final tetrad 
\begin{align}
    \vec{l} &= \vec{\p_v}+U\vec{\p_s}+X^\Theta\vec{\p_\Theta}+X^{\tilde{\Phi}}\vec{\p_{\tilde{\Phi}}}\,, \\
    \vec{n} &= \vec{\p_s}\,,\\
    \vec{\dot{m}} &= \Omega\vec{\p_s}+\xi^\Theta\vec{\p_\Theta}+\xi^{\tilde{\Phi}}\vec{\p_{\tilde{\Phi}}}\,,
\end{align}
where the functions $U,X^\Theta,X^{\tilde{\Phi}},\Omega,\xi^\theta,\xi^{\tilde{\Phi}}$ are given as follows
\begin{widetext}
\begin{align}
U(r,\Theta) &=  E(r,\theta)\bar{E}(r,\theta)+ 
\cos^2\Theta+\frac{\Sigma^-(r,\theta)}{2\ka^2}+
\left[
\frac{iE(r,\theta)}{\sqrt{2}\bar{\vrho}(r,\theta)}
\left(\frac{i\Sigma_0(r)\Sigma(r,\Theta)}{\ka\sqrt{\UpsPi(r)}}-\ka\left(\cos^2\Theta-\cos^2\theta\right)\tan\theta\right)
+\mathrm{c.c.}\right],\\
X^\Theta(r,\Theta) &= \left\{
\frac{1}{\ka}+
\left[ \frac{E(r,\theta)}{\sqrt{2}\bar{\vrho}(r,\theta)}\left(\frac{\Sigma_0(r)}{\sqrt{\UpsPi(r)}}+i\tan\theta\right)+\mathrm{c.c.}\right]\right\}\cos\Theta,\\
X^{\tilde{\Phi}}(r,\Theta) &= -\frac{\Rp}{2M\ka}-\left[
\frac{E(r,\theta)}{\sqrt{2}\bar{\vrho}(r,\theta)}\left(\frac{r^2}{\sqrt{\UpsPi(r)}}+\cot\theta\right)+\mathrm{c.c.}\right],\\
\Omega(r,\Theta) &=\bar{E}(r,\theta)+\frac{1}{\sqrt{2}\bar{\vrho}(r,\theta)}
\left(\frac{i\Sigma_0(r)\Sigma(r,\Theta)}{\ka\sqrt{\UpsPi(r)}}-\ka\left(\cos^2\Theta-\cos^2\theta\right)\tan\theta\right),
\end{align}
\end{widetext}
\begin{align}
\xi^\Theta(r,\Theta) &= \frac{i\cos\Theta}{\sqrt{2}\bar{\vrho}(r,\theta)}\left(\frac{\Sigma_0(r)}{\sqrt{\UpsPi(r)}}+i\tan\theta\right),\\
\xi^{\tilde{\Phi}}(r,\Theta)&=\frac{-1}{\sqrt{2}\bar{\vrho}(r,\theta)}\left(\frac{i\, r^2}{\sqrt{\UpsPi(r)}}+\cot\theta\right),
\end{align}
where we implicitly assume $r=r(s,\Theta)$ and also $\theta=\theta(r,\Theta)=\theta(r(s,\Theta),\Theta)$. We also introduced
\begin{equation}
    \Sigma^-(r,\theta)=\frac{1}{2} \,\vec{f}_{ab}\vec{f}^{ab} =r^2-\ka^2\cos^2\theta\,.
\end{equation}

The following  identities were used during the simplification process
\begin{align}
    \frac{\p}{\p s}\, r(s,\Theta) &= \frac{\sqrt{\UpsPi(r)}}{\Sigma(r,\theta(r,\Theta))} \,, \\
    \frac{\p}{\p r}\,  s(r,\Theta) &=\frac{\Sigma(r,\theta(r,\Theta))}{\sqrt{\UpsPi(r)}} \,, \\
    \frac{\p}{\p \Theta}\, s(r,\Theta) &= \ka\, \frac{\cos^2\Theta-\cos^2\theta(r,\Theta)}{\cos\Theta} \,,\\
    \frac{\p}{\p \Theta}\, r(s,\Theta) &= -\ka\,\frac{\sqrt{\UpsPi(r)}}{\Sigma(r,\theta(r,\Theta))}\frac{\cos^2\Theta-\cos^2\theta(r,\Theta)}{\cos\Theta} \,, \\
    \frac{\p}{\p\Theta}\, \theta(r,\Theta) &= \frac{\cos\theta(r,\Theta)}{\cos\Theta} \,.
\end{align}

Also, we can simplify the form of the null rotation $E$, given in Eq.~(\ref{eq:rotE}), as
\begin{equation}
    E(r,\theta) = \frac{1}{\sqrt{2}}\left(-e^{-i\Theta}+i\,\frac{\mathcal{I}_2(r)+\Rp-\ka\sin\theta}{\ka}\right)\,.
\end{equation}
The null rotation parameter $E$ has the following relation to $s$
\begin{align}
    E(r,\Theta)&=i\,\frac{s(r,\Theta)+\Rp+i\ka\cos\Theta}{\sqrt{2}\,\ka} \,.
\end{align}

In the coordinates $(v,s,\Theta,\tilde{\Phi})$ the metric takes the following form
\begin{equation}
    \vec{g}_{\mu\nu} = 
    \begin{pmatrix}
    g_{vv} & 1 & g_{v\Theta} & g_{v\tilde{\Phi}} \\
    1 & 0 & 0 & 0 \\
    g_{\Theta v} & 0 & g_{\Theta\Theta} & g_{\Theta\tilde{\Phi}} \\
    g_{\Phi v} & 0 & g_{\tilde{\Phi}\Theta} & g_{\tilde{\Phi}\tilde{\Phi}}
    \end{pmatrix} ,
\end{equation}
where 
\begin{align}
g_{vv} &= \frac{4M^2\Rp\,\Delta-\Rm\sin^2\theta\left(4M^2\Rp(\Rp-2r)+\Upsilon(r,\theta)\right)}{4M^2\Rp\,\Sigma(r,\theta)} \,,\\
g_{v\Theta} &= \ka\,\frac{\cos^2\theta}{\cos\Theta}\left(\frac{r}{\Rp}\frac{\Sigma(\Rp,\theta)}{\Sigma(r,\theta)}-\frac{\cos^2\Theta}{\cos^2\theta}\right) \,, \\
g_{v\tilde{\Phi}} &= -\frac{\Rm}{2M\ka}\frac{4M^2\Rp\, r-\Upsilon(r,\theta)}{\Sigma(r,\theta)}\,\sin^2\theta \,, \\
g_{\Theta\Theta} &= -\frac{\cos^2\theta}{\cos^2\Theta}\left(r^2+\frac{2M\ka^2r\cos^2\theta}{\Sigma(r,\theta)}\right)\,, \\
g_{\Theta\tilde{\Phi}} &= \frac{2M\ka^2 r }{\Sigma(r,\theta)}\frac{\cos^2\theta}{\cos\Theta}\,\sin^2\theta\,, \\
g_{\tilde{\Phi}\tilde{\Phi}} &= -\frac{\Upsilon(r,\theta)}{\Sigma(r,\theta)}\,\sin^2\theta\,.
\end{align}

As has been discussed in a paragraph after the Eq.~(\ref{eq:sint}), the inversion of the explicit relation $s=s(r,\Theta)$ is not analytically possible. Therefore, also the above stated metric functions contains implicitly defined coordinates.

\section{Uniqueness}
\label{sec:uni}

First of all we notice that for an arbitrary choice of constants $E,K,\epsilon_r,\epsilon_\theta$ (the defining parameters of the vector field $\vec{n}$), we can complete the tetrad in such a way, that $\vec{l},\vec{m},\vec{\bar{m}}$ are tangent to the horizon (on the horizon) and start evolving from the horizon. But these coordinate systems suffer the same disadvantages (and, actually, are even ``worse'': not defined everywhere, not having mirror symmetry with respect to the equatorial).

In order to find a field which is defined everywhere and is regular around the axis we need to investigate the twist-free geodesic congruence in a more general setting.

The vector field $\vec{u}$ given by Eq.~(\ref{eq:genu}) is a geodesic congruence if
\begin{align}
    \vec{u}^a\nabla_a E(r,\theta) &= 0\,, &
    \vec{u}^a\nabla_a L(r,\theta) &= 0\,, &
    \vec{u}^a\nabla_a K(r,\theta) &= 0\,.
\end{align}
Specializing for null geodesic, the twist-free condition 
\begin{align}
    \vec{n}_{[a}\nabla_{b}\vec{n}_{c]} = 0
\end{align}
leads to 
\begin{align}
    L = \mathrm{const} \times E\,,
\end{align}
and subsequently thus vanishing angular momentum $L=0$, since we need $\vec{n}$ to be defined close to the axis.

Then the energy $E$ can be scaled to a constant and we need to solve the equation for the Carter's constant
\begin{align}
    \frac{\epsilon_r K_{,r}}{\sqrt{E\SigZ^2-K\Delta}}+\frac{\epsilon_\theta K_{,\theta}}{\sqrt{K-\ka^2E^2\sin^2\theta}} = 0\,.
    \label{eq:K}
\end{align}
If this equation is satisfied, $\vec{n}$ is geodesic and twist-free.
So far the only explicitly known solutions to this partial differential equation are $K=\mathrm{const}$ and we need to overcome this complication. Notice that
$$K(r,\theta)=\ka^2 E^2\sin^2\theta $$ 
everywhere does not lead to a geodetic vector field.

Let us define the initial values for $\vec{n}$ on the horizon and then evolve the geodesic equations.
A natural choice for the Carter's constant on the horizon is 
\begin{align}
    K &\doteq a^2E^2\sin\theta^2\,,
\end{align}
since it makes the $\theta$ component of the field $\vec{n}$ vanishing.

In the following we need to work in null coordinates given by 
\begin{align}
    t&=u-\int\frac{\ka^2+r^2}{\Delta(r)}\,\d r\,, \\
    \phi&=\tilde{\phi}-\int\frac{\ka}{\Delta(r)}\,\d r\,.
\end{align}

Introducing
\begin{align}
    R(r,\theta) &= E^2\SigZ^2(r)-K(r,\theta)\Delta(r)\,, \label{eq:Rf}\\
    S(r,\theta) &= K(r,\theta)-\ka^2E^2\sin^2\theta\,, \label{eq:Sf}
\end{align}
we thus look for an integral lines of the field 
\begin{align}
    \vec{n} &= \frac{1}{\Sigma} \Biggl( \frac{\SigZ\sqrt{R}+E\left(\SigZ^2-\ka^2\sin^2\theta\, \Delta\right)}{\Delta}\,\vec{\p_u} \nonumber\\
    &\qquad+\epsilon_r\sqrt{R}\,\vec{\p_r}+\epsilon_\theta\sqrt{S}\,\vec{\p_\theta} +\ka\frac{2MEr+\epsilon_r\sqrt{R}}{\Delta}\,\vec{\p_{\tilde{\phi}}} \Biggr) 
\end{align}
with the initial values for the  vector field $\vec{n}$ on the horizon
\begin{align}
    \vec{n}&\doteq \frac{1}{2\Sigma(\Rp,\theta_0)} \Biggl(\ka^2\sin^2\theta_0\,\vec{\p_u}+2M\Rp\,\vec{\p_r} \nonumber \\
    &\qquad {+\;\color{lightgray}0\, \vec{\p_\theta}}  +\left(\ka+\frac{\Rm \Sigma(\Rp,\theta_0)}{2\ka M}\right)\vec{\p_{\tilde{\phi}}}\Biggr),
    \label{eq:inic}
\end{align}
where we already set $E=-1$ and $\epsilon_r=1$. We shall see later that there is no dependence on $\epsilon_\theta$.

Using the recent developments in analytical treating of geodesics within the Kerr spacetime \cite{hackmann_analytical_2014,cieslik_kerr_2023} (and following the notation of \cite{cieslik_kerr_2023}) we can write the explicit solution in terms of the Weierstrass $\wp$ function as
\begin{widetext}
\begin{align}
r(z) &= r_0+\frac{-\eps_r\sqrt{R(r_0)}\,\wp_r'(z)+\frac{1}{2}R'(r_0)\left(\wp_r(z)-\frac{1}{24}\,R''(r_0)\right)+\frac{1}{24}R(r_0)R'''(r_0)}{2\left(\wp_r(z)-\frac{1}{24}R''(r_0)\right)^2-\frac{1}{48}R(r_0)R^{(4)}(r_0)}\,, \label{eq:geoa}  \\
\mu(z) &=  \mu_0+\frac{\eps_\theta\sqrt{G(\mu_0)}\,\wp_\theta'(z)+\frac{1}{2}G'(\mu_0)\left(\wp_\theta(z)-\frac{1}{24}\,G''(\mu_0)\right)+\frac{1}{24}G(\mu_0)G'''(\mu_0)}{2\left(\wp_\theta(z)-\frac{1}{24}G''(\mu_0)\right)^2-\frac{1}{48}G(\mu_0)G^{(4)}(\mu_0)} \,, \label{eq:geos}
\end{align}
\end{widetext}
where $z$ is a Mino time and the angular variable has been re-parameterize such that $\mu=\cos\theta$. Therefore, also $G$ is re-parameterized $S$ from Eq.(\ref{eq:Sf}) and is of the form
\begin{align}
    G(\mu) &= (1-\mu^2)\left[K-\ka^2(1-\mu^2)\right]\,.
\end{align}
The complete form of the Weierstrass functions $\wp_r(z)$ and $\wp_\theta(z)$ where the Weierstrass invariants $g_2$, $g_3$ (which are given in terms of roots of $R$ or $G$) is as follows
\begin{align}
    \wp_r(z) &\equiv \wp(z;g^r_2,g^r_3)\,,\\
    \wp_\theta(z) &\equiv \wp(z;g^\theta_2,g^\theta_3)\,.
\end{align}
The Weierstrass invariants expressed in terms of black hole parameters and the Carter constant read
\begin{align}
    g^\theta_2 &= \frac{1}{12}\left(K^2-16\ka K+16\ka^2\right),\\
    g_3^\theta &= \frac{1}{216}\left(K-2\ka\right)\left(K^2+32\ka K-32\ka^2\right)\,,\\
    g_2^r &= g^\theta_2\,, \\
    g^r_3 &=g_3^\theta-\frac{K^2M^2}{4}\,,
\end{align}

Let us emphasize, that $K(r,\theta)$ varies as a field over the manifold, but it is constant along every single geodesic.

In the Eqs.~(\ref{eq:geoa})\,--\,(\ref{eq:geos}) we will start from the horizon, thus $r_0=\Rp$, and we will consider $\theta_0$, being constant along the geodesic as the new coordinate $\Theta$. Second new coordinate will be the Mino time $z$. Taking the initial conditions into the account, the expression for $\mu(z)$ greatly simplifies and we can explicitly write $\theta(z)=\arccos\mu(z)$. Thus we can turn a solution to the geodesic equation into the coordinate transformation as follows 
\begin{align}
    r(z,\Theta) &= \text{Eq.~(\ref{eq:geoa}) with }r_r=\Rp \label{eq:Kr} \,,\\
    \theta(z,\Theta) &= \arccos\left( \cos\Theta + \frac{6\ka^2\sin\Theta\sin 2\Theta}{\ka^2(3+5\cos^2\Theta)+24\wp_\theta(z)}\right)\,. \label{eq:Kt}
\end{align}

\begin{figure}
    \centering
    \begin{overpic}[keepaspectratio,width=.2\textwidth,percent]{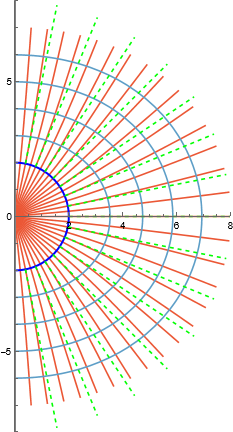}
    \put(55,42){$R$}
    \put(-2,95){$z$}
    \end{overpic}
    \caption{The integral lines of $\vec{n}$ and equipotentials of affine parameter in BL coordinates for $\epsilon_r=1$ emanating from the horizon with initial conditions given by Eq.~(\ref{eq:inic}). The geodesics, i.e. lines of constant $\Theta$, are slightly bent which can be difficult to observe by naked eye, therefore we included lines of constant $\theta$ as green dashed lines (with half density of lines). The explanation of structures is the same as in Fig.~\ref{fig:coords}.}
    \label{fig:Kcoord}
\end{figure}

\begin{figure}
    \centering
    \begin{overpic}[keepaspectratio,width=.45\textwidth,percent]{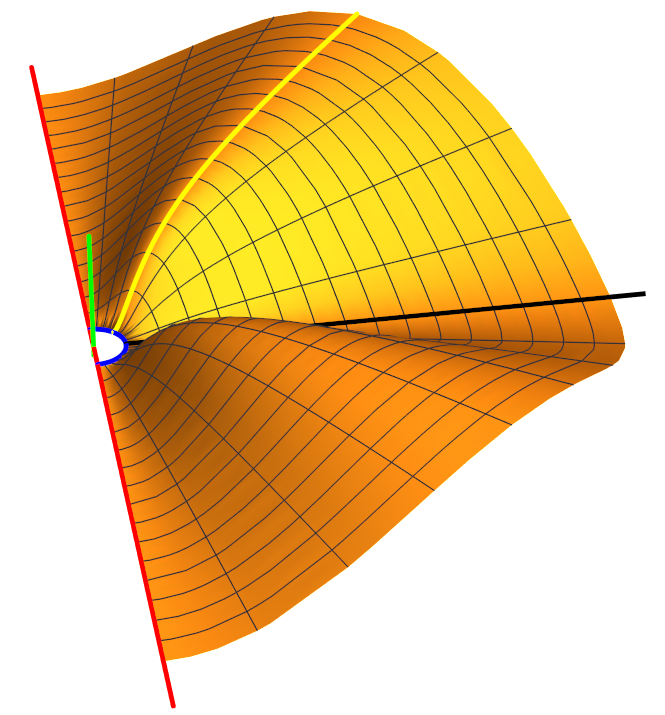}
    \put(2.5,93){$z$}
    \put(90,54){$R$}
    \put(2,62){$\Delta K$}
    \end{overpic}
    \caption{The value of $\Delta K\equiv K(r(z,\Theta),\theta(z,\theta))-\ka^2\sin^2\theta$ above the horizon (small half-circle in the middle) in BL coordinates. The red line is a rotational axis ($z$), the black line depicts the equatorial plane, in the green direction the function value is plotted. The exact values have no particular meaning. The yellow line is one particular geodesic with $\Theta=\pi/4$ for which the function is in greater detail in Fig.~\ref{fig:cpi}. The blue line is a black hole horizon.}
    \label{fig:Kval}
\end{figure}

Now it is straightforward to perform the change of variables given by coordinate transformation (\ref{eq:Kr})\,--\,(\ref{eq:Kt}) in the partial differential equation for the Carter's constant $K$ given by Eq.~(\ref{eq:K}) and check directly that 
\begin{align}
    K(z,\Theta) &= \ka^2\sin^2\Theta\,,
\end{align}
is a valid solution. The value of $K$ differs just slightly from the $\ka^2\sin^2\theta$. In the Fig.~\ref{fig:Kval} we plot the value of expression 
$$K(r(z,\Theta),\theta(z,\theta))-\ka^2\sin^2\theta$$ 
to visualize these small differences. We can see that there is no difference neither on the axes nor the equatorial.

\begin{figure}
    \centering
    \begin{overpic}[keepaspectratio,width=.45\textwidth,percent]{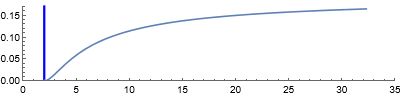}
    \put(96,7){$r$}
    \put(-2,25){$\Delta K$}
    \end{overpic}
    \caption{The vertical axis represents the value of $\Delta K\equiv K(r(z,\Theta),\theta(z,\theta))-\ka^2\sin^2\theta$ for one particular geodesic from Fig.~\ref{fig:Kval}. The values of the black hole are $\Rp=2,\,\Rm=1$ and $\Theta=\pi/4$. On the horizontal axis is the BL coordinate $r$.}
    \label{fig:cpi}
\end{figure}

The coordinates are constructed such that 
\begin{itemize}
    \item on the horizon we have $\Theta\doteq\theta$,
    \item the axes and equatorial plane are generated by geodesics,
    \item the expansion is positive and bounded (the explicit formula is exceedingly long for presentation),
    \item the vector field $\vec{n}$ is smooth.
\end{itemize} 

The Mino time $z$ is given by 
\begin{align}
    z(r) &= \epsilon_r \int_{\Rp}^r \frac{1}{\sqrt{R(r')}}\,\d r'.
\end{align}
The geodesic leaving horizon at $z=0$ reach infinity in finite Mino time $z_\infty$ which can be evaluated in terms of elliptic integrals as follows
\begin{align}
    z_\infty &= \left[\frac{2F\left(\phi_\infty,\frac{(r_2-r_3)(r_1-r_4)}{(r_1-r_3)(r_2-4_4)}\right)}{\sqrt{(r_1-r_3)(r_2-r_4)}}\right]_{r=\Rp}^{r=\infty}\,,\\
    \phi_\infty &= \arcsin\left(\frac{\sqrt{r-r_1}\sqrt{r_2-r_4}}{\sqrt{r-r_2}\sqrt{r_2-r_4}}\right)\,.
\end{align}
The detailed factorization of polynomial $R(r)$ is given in Appendix~\ref{app:Rf}.

In principle it may be feasible to write down the complete solution of the geodesic equations 
\begin{align}
    t &= t(t_0,z,\Theta) \,,\\
    r &= r(z,\Theta) \,,\\
    \theta &= \theta(z,\Theta) \,,\\
    \phi &= \phi(\phi_0,z,\Theta)\,,
\end{align}
in terms of the Weierstrass $\wp$ functions, see \cite{cieslik_kerr_2023}.
Alas, to proceed further with the explicit coordinate transformation to coordinates $t_0,z,\Theta,\phi_0$ and solving the parallel transport equations seems to be too complicated.

But we advocate that this vector field $\vec{n}$ is the best one for construction of a null foliation of the Kerr spacetime generated by non-twisting congruence. It is, in some sense, unique, since it does not depend on the choice of $\epsilon_\theta$.

\section{Comparison with Bondi coordinates}
\label{sec:bondi}
The coordinates $(v,s,\Theta,\Phi)$ were called ``Bond-like'' in \cite{scholtz_kerr-newman_2017}. It's difficult to judge whether such a designation is correct or not since the rigorous definition of the ``likeness'' has not been provided.

In the original Bondi\,--\,Sachs formalism the coordinate system is based upon a congruence of null geodesics emanating from a spatial spherical cross section of a chosen world-tube. The null coordinate labels the cross sections and the angular coordinates  are the introduced on such a cross section (and stay constant along the rays). The radial distance is then a parameter along a particular ray. The metric is then of the form
\begin{multline}
    \d s^2 = -\frac{V}{r}\,e^{2\beta}\d u^2-2e^{2\beta}\,\d u\,\d r \\+r^2 h_{AB}\left(\d x^A-U^A\d u\right)\left(\d x^B-U^B\d u\right),
\end{multline}
with $h_{AB}$ being the metric on topological spheres, such that $\det h_{AB}=\det q_{AB}$, where $q_{AB}$ is metric on a sphere. The coordinate $r$ is called a luminosity distance since a surface $u=\mathrm{const}$ and $r=\mathrm{const}$ has a surface $4\pi r^2$.

A version of Bondi\,--\,Sachs approach, which fits more appropriately to NP formalism, is affine-null metric formulation provided by Sachs and Winicour \cite{winicour_affine-null_2013}, where the affine parameter along the null rays is used instead of the luminosity distance.
A simple transformation 
\begin{equation}
    \p_{r}\lambda(u,r,x^A) = e^{2\beta}\,,
\end{equation}
leads to the metric in the form
\begin{multline}
    \d s^2 = -\left(\mathcal{V}-g_{AB}W^AW^B\right)\d u^2-2\d u\d \lambda \\ 
    -2g_{AB}W^A\,\d u\,\d x^B+g_{AB}\d x^A\d x^B\,,
\end{multline}
with $g_{AB}=r^2h_{AB}$ and $\det h_{AB}=\det q_{AB}$, where $q_{AB}$ is metric on a round sphere. Now, $r,\,\mathcal{V},\,W^A$ as well as $h_{AB}$ are functions of $(u,\lambda,x^A)$.

The null geodetic congruence is given by
\begin{align}
    \vec{n}_W &= \vec{\p_\lambda}\,,
\end{align}
and the expansion of this vector field
\begin{align}
    \Theta_{\vec{n}_W} &= \nabla_a\vec{n}_W^a = -\frac{1}{2}\frac{\p \det g_{AB}}{\p \lambda} = 4 r^3 r_{,\lambda}\,.
\end{align}

Clearly, the coordinate system of formalism of isolated horizon is a member of affine-null metric formulation. It is moreover endowed with a particular parallely propagated null tetrad.

\section{Conclusions}
\label{sec:con}
We explicitly constructed a parallely propagated null tetrad and an appropriate coordinate system of Kerr metric, following the existing work on the formulation of Kerr solution in the formalism of isolated horizons. We provided a physical  interpretation and showed that these coordinate systems are not well behaved (not covering the whole space-time, the vector field $\vec{n}$ being irregular on the axis with diverging expansion). 

Demanding the regularity of the vector field $\vec{n}$ we found a well behaved geodetic congruence in terms of the Weierstrass $\wp$ functions. We suggest that this is the proper non-twisting null geodetic field. Alas, the coordinate transformation is extremely difficult to treat analytically and we were able neither to cast the Kerr metric in coordinates adapted to this congruence neither to construct a parallely propagated tetrad in the sense of isolated horizons.

\begin{acknowledgements}
    D.K. acknowledges the support from the Czech Science Foundation, Grant 21-11268S. The calculations were performed using \emph{Wolfram Mathematica}$^{\circledR}$ and the powerful package \emph{xAct}.

\end{acknowledgements}

\appendix

\section{NP quantities of Kerr black hole}
\label{app:Kerr}
The nonzero NP spin coefficients corresponding to the tetrad (\ref{eq:NPtetrad}) are 
\begin{equation}
\begin{alignedat}{2}
\tilde{\pi} &= \frac{i}{\sqrt{2}}\,\frac{\ka \sin\theta}{\Krho^2}\,,\qquad &
\tilde{\mu} &= \frac{-1}{\sqrt{2}}\,\frac{\Delta}{\Sigma\Krho}\,, \\
\tilde{\tau} &= \frac{-i}{\sqrt{2}}\,\frac{\ka \sin\theta}{\Sigma}\,, &
\tilde{\rho} &= \frac{-1}{\sqrt{2}}\,\frac{1}{\Krho}\,,\\
\tilde{\gamma} &=\tilde{\mu} +\frac{1}{\sqrt{2}}\,\frac{r-M}{\Sigma}\,,\qquad &
\tilde{\beta} &= \frac{1}{\sqrt{2}}\,\frac{\cot \theta}{\Krhocc}\,, \\
\tilde{\alpha} &= \tilde{\pi}-\bar{\tilde{\beta}}\,, & 
\end{alignedat}\label{eq:spinc}
\end{equation}
and the only nonzero Weyl scalar reads 
\begin{equation}
\tilde{\psi}_2 = -\frac{M}{\Krho^3} \,.
\label{eq:psi2}\end{equation}

\section{The explicit form of integral $\mathcal{I}_1$ and $\mathcal{I}_2$}
\label{app:eli}
The integrals $\mathcal{I}_0$ and $\mathcal{I}_2$ have been introduced in Eqs.~(\ref{eq:I0})\,--\,(\ref{eq:I2}) as
\begin{align}
    \mathcal{I}_0(r) &= \int_{r_p}^r\frac{1}{\sqrt{\UpsPi(u)}}\,\d u \,, \\
    \mathcal{I}_2(r) &= \int_{r_p}^r\frac{u^2}{\sqrt{\UpsPi(u)}}\,\d u \,.
\end{align}
In order to evaluate these (in terms of elliptic integral) we need to factorize the function $\UpsPi(u)$ as
\begin{align}
    \UpsPi(u) & = u(u^3+\Rp\Rm u+\Rp\Rm(\Rp+\Rm)) \nonumber\\
    &=u(u^3+\ka^2 u+2M\ka^2) \nonumber\\
    &=u(u-u_r)(u-u_c)(u-\bar{u_c}) \nonumber\\
    &=u(u+2Z)(u^2-2Zu+R)
\end{align}
which has two real roots $0$, $u_r$ and two mutually complex conjugated roots $u_c$ and $\bar{u_c}$, namely
\begin{align}
    u_r &= -\frac{2^{1/3}\ka^2}{3^{2/3}w}+\frac{w}{2^{1/3}3^{2/3}}\,,\\
    u_c &= \frac{(1+i\sqrt{3})\ka^2}{2^{2/3}3^{1/3}w}-\frac{(1-i\sqrt{3})w}{2^{4/3}3^{1/3}w}\,,
\end{align}
where
\begin{align}
    w &= 2\ka^2(\sqrt{3}\sqrt{27M^2+\ka^2}-9M)\,,\\
    Z &= (u_c+\bar{u_c})/2 = -u_r/2 \,,\\
    R &= \sqrt{u_c \bar{u_c}}\,.
\end{align}

Then we can write the integrals $\mathcal{I}_0$ and $\mathcal{I}_2$ easily in terms of primitive functions
\begin{align}
    \tilde{\mathcal{I}}_0(u)&=2i\frac{1}{\sqrt{q}}F\left(\phi_0|m_0\right)\,,\\
    \phi_0 &= \arcsin\left(\frac{1}{\sqrt{2}}\sqrt{1+\frac{Z-R^2/u}{p}}\right)\,,\\
    m_0 &= \frac{4Zp}{q}\,,
\end{align}
and
\begin{align}
    \tilde{\mathcal{I}}_2(u)&=\sqrt{-Q} E\left(\phi_2|m_2\right)+\sqrt{-\frac{R^4}{Q}} F\left(\phi_2|m_2\right) \nonumber \\
    &\quad+\sqrt{\frac{u(u^2-2uZ+R^2)}{r+2Z}}\,,\\
    \phi_2 &= \arcsin\left(\frac{1}{\sqrt{2}}\sqrt{1-\frac{Z+\frac{R^2-4Zu}{u+2Z}}{p}}\right),\\
    m_2 &= -\frac{4Zp}{Q}\,,
\end{align}
with constants
\begin{align}
    p &= \sqrt{Z^2-R^2}\,,\\
    q &= R^2+2Zp+2Z^2\,,\\
    Q &= R^2-2Zp+2Z^2\,.
\end{align}
To clarify the notation we use the incomplete elliptic integral of the first kind $F$ and of the second kind $E$ which are defined as
\begin{align}
    F(\phi|m) &= \int_0^\phi\frac{1}{\sqrt{1-m\sin^2\theta}}\,\d \theta\,,\\
    E(\phi|m) &= \int_0^\phi\sqrt{1-m\sin^2\theta}\,\d \theta\,.
\end{align}

\section{The choice $\epsilon_\theta=1$}
\label{sec:differentEps}
\begin{figure}
    \centering
    \begin{overpic}[keepaspectratio,width=.2\textwidth,percent]{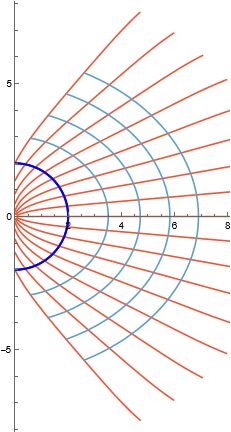}
    \put(55,42){$R$}
    \put(-2,95){$z$}
    \end{overpic}
    \caption{The integral lines of $\vec{n}$ and equipotentials of affine parameter in BL coordinates for $\epsilon_r=\epsilon_\theta=1$ emanating from the horizon. The detailed explanation as in Fig.~\ref{fig:coords}.}
    \label{fig:c2}
\end{figure}

The other choice of $\epsilon_r=\epsilon_\theta=1$ was investigated in \cite{scholtz_kerr-newman_2017}. The main disadvantage of this choice is that the appropriate coordinates does not cover a huge portion of space-time outside the horizon as can be seen in Fig.~\ref{fig:c2}. On the other hand the expansion of this congruence newer changes the sign, yet it is still divergent in the vicinity of the axis. The explicit value is
\begin{equation}
    \Theta_{\vec{n}} = \nabla^a\vec{n}_a = \frac{1}{\Sigma}\left(\frac{\d \UpsPi(r)}{\d r}+\ka\,\frac{\cos 2\theta}{\cos\theta} \right) ,
\end{equation}

The coordinate transformations given by Eqs.~(\ref{eq:Theta}\,--\,\ref{eq:Phi}) and (\ref{eq:theta}\,--\,\ref{eq:phi}) differ just by different signs in relations for $\phi$ and $\Phi$ and interchanging $\Gamma^- \longleftrightarrow \Gamma^+$.

In \cite{fletcher_kerr_2003} they were propagating the null geodesics from null infinity, thus covering the whole space-time.

\section{Factorization of polynomial $R(r)$}
\label{app:Rf}
In this Appendix we provide the factorization of polynomial $R(r)$ which has been introduced in Eq.~(\ref{eq:Rf}). Taking into the account that we set $E=-1$ we need to factorize
\begin{align}
    R &= \SigZ^2-K\Delta(r) \nonumber\\
     &= (r^2+\ka^2)^2-\ka^2(r-\Rp)(r-\Rm)\sin^2\Theta \nonumber\\
     &=(r-r_1)(r-r_2)(r-r_3)(r-r_4)\,.
\end{align}
This can be done in terms of real variables $r_r,r_i^\pm$ as follows
\begin{align}
    r_1 &= r_r+ir_i^+ \,, & r_3 &= -r_r+ir_i^- \,,\\
    r_2 &= r_r-ir_i^+ \,, & r_3 &= -r_r-ir_i^- \,,
\end{align}
where
\begin{align}
    r_r &= \frac{1}{2\sqrt{6}}\Biggl[2^{2/3}(q+\sqrt{w})^{1/3}+24 g_2^\theta \left(\frac{2}{q+\sqrt{w}}\right)^{1/3} \nonumber \\
    &\qquad\qquad-2\ka^2(3+\cos 2\Theta)\Biggr]^{1/3}\,,\\
    r_i^\pm &= \frac{\mp 1}{2}\sqrt{4r_r^2+\ka^2(3+2\cos 2\Theta)\pm\frac{2M\ka^2\,\sin^2\Theta}{r_r}}\,,\\
    g_2^\theta &=\frac{1}{96}\ka^4\left(67+60\cos2\Theta+\cos4\Theta\right)\,,\\
    q &= \ka^2\biggl[-128\ka^2+192\ka^2\sin^2\Theta \nonumber \\
    &\qquad\qquad+12(9M^2-5\ka^2)-2\ka^2\sin^6\Theta\biggr]\,,\\
    w &= q^2-6912 (g_2^\theta)^2\,.
\end{align}
\bibliography{kofron}

\end{document}